\documentclass[twocolumn,showpacs,aps,prd,amsfonts,
superscriptaddress,nofootinbib]{revtex4}
\usepackage{epsfig}
\usepackage{bm}

\begin{document}

\newcommand*{\nl}{\nonumber \\}
\newcommand*{\bea}{\begin{eqnarray}}
\newcommand*{\eea}{\end{eqnarray}}
\newcommand*{\bi}{\bibitem}
\newcommand*{\be}{\begin{equation}}
\newcommand*{\ee}{\end{equation}}
\newcommand*{\rms}{M_\rho^2(s)}
\newcommand*{\mrs}{m_\rho^2}
\newcommand*{\ra}{\rightarrow}
\newcommand*{\die}{e^+e^-}
\newcommand*{\eepppp}{e^+e^-\rightarrow4\pi}
\newcommand*{\eeppppn}{e^+e^-\rightarrow\pi^+\pi^-\pi^0\pi^0}
\newcommand*{\eeppppc}{e^+e^-\rightarrow\pi^+\pi^-\pi^+\pi^-}
\newcommand*{\rppppn}{\rho^0\rightarrow\pi^+\pi^-\pi^0\pi^0}
\newcommand*{\rppppc}{\rho^0\rightarrow\pi^+\pi^-\pi^+\pi^-}
\newcommand*{\pmnn}{\pi^+\pi^-\pi^0\pi^0}
\newcommand*{\pmpm}{\pi^+\pi^-\pi^+\pi^-}
\newcommand*{\arp}{{a_1\rho\pi}}
\newcommand*{\hrp}{{h_1\rho\pi}}
\newcommand*{\orp}{{\omega\rho\pi}}
\newcommand*{\eg}{e.g.}
\newcommand*{\rp}{{\rho^\prime}}
\newcommand*{\bp}{{\beta^\prime}}
\newcommand*{\dep}{{\delta^\prime}}
\newcommand*{\ep}{{\epsilon^\prime}}
\newcommand*{\rpp}{{\rho^{\prime\prime}}}
\newcommand*{\amu}{{\mathbf A}^\mu}
\newcommand*{\anu}{{\mathbf A}^\nu}
\newcommand*{\amunu}{{\mathbf A}^{\mu\nu}}
\newcommand*{\vmu}{{\mathbf V}_\mu}
\newcommand*{\vmunu}{{\mathbf V}_{\mu\nu}}
\newcommand*{\fai}{\mathbf\phi}
\newcommand*{\rf}[1]{(\ref{#1})}
\newcommand*{\mas}{m_{a_1}^2}
\newcommand*{\lag}{{\mathcal L}}
\newcommand*{\opava}{Institute of Physics, Silesian University in Opava,
Bezru\v{c}ovo n\'{a}m\v{e}st\'{i} 13, 746 01 Opava, Czech Republic}
\newcommand*{\praha}{Institute of Experimental and Applied Physics,
Czech Technical University, Horsk\'{a} 3/a, 120 00 Prague, Czech
Republic}

\title{Joint description of the \bm{$\die$} annihilation into both
four-pion channels}

\author{Josef Jur\'{a}\v{n}}
\affiliation{\opava}
\author{Peter Lichard}
\affiliation{\opava}
\affiliation{\praha}

\date{\today}

\begin{abstract}
The $\eeppppn$ reaction cross section as a function of the incident
energy is calculated using a model that is an extension of our recently
published model of the $\die$ annihilation into four charged pions. The
latter considered the intermediate states with the $\pi$, $\rho$, and
$a_1$ mesons and fixed the mixing angle of the $\arp$ Lagrangian and
other parameters by fitting the cross section data. Here we supplement
the original intermediate states with those containing $\omega(782)$
and $h_1(1170)$, but keep unchanged the values of those parameters that
enter both charged and mixed channel calculations. The inclusion of
$\omega$ is vital for obtaining a good fit to the cross section data,
while the intermediate states with $h_1$ further improve it. Finally,
we merge our models of the $\eeppppn$ and $\eeppppc$ reactions and
obtain a simultaneous good fit.
\end{abstract}

\maketitle

The electron-positron annihilation into four pions has been theoretically
studied by several authors \cite{renard,decker,achasov12,czyz,ecker34}.
Assuming the one-photon approximation
and vector meson dominance, this process goes via the $\rho$(770) meson
and its recurrences. If some conditions on the rho-decay amplitude into
four pions are met \cite{app}, the cross section of the $\die$
annihilation into four pions at invariant energy $W$ can be expressed
in terms of the decay width of a $\rho$ meson with mass $W$ into four
pions \cite{achasov12,licjur}. Some of the models of the four-pion
decay of the $\rho$ meson \cite{bramon,eidelman,plant,achasov12} can
thus be conveniently utilized when determining the excitation function
of the $\die$ annihilation into four pions.

In our recent work \cite{licjur} we calculated the excitation function
of the $\die$ annihilation into four charged pions and compared it to
the existing data. We confirmed the conclusion of several experimental
and theoretical papers \cite{decker,czyz,ecker34,achasov3,cmd2} that
the axial-vector isovector resonance $a_1(1260)$ plays an important
role. The new feature of our approach was that we did not take some
\textit{a priori} chosen Lagrangian of the $\arp$ interaction, but
considered a two-component Lagrangian that contained two parameters: a
mixing angle and an overall coupling constant. We varied the mixing
angle in an effort to get the best fit to the data. The coupling
constant was determined for each mixing angle from the given total
width of the $a_1$ resonance. Besides the intermediate states with the
$a_1$ resonance, we considered also those with only pions and $\rho$'s
as given in various theoretical schemes
\cite{bramon,eidelman,plant,achasov12}. They influence the calculated
cross section mainly in the rho mass region. However, the quality of
the fit throughout the energy region covered by BaBar experiment
\cite{babar} has improved only slightly.

As far as the $\die$ annihilation cross section data are concerned, the
situation in the $\pmnn$ sector is worse than in the $\pmpm$ one. The
data coming from various experimental groups did not agree with one
another very well; see, \eg, Fig.~10 in \cite{druzhinin}. It is good
news that the latest published data by the SND Collaboration at BINP in
Novosibirsk \cite{snd} agree well with the newest BaBar
\cite{druzhinin,petzold} and CMD-2 \cite{druzhinin,logashenko} data.
Unfortunately, those data are still preliminary and publicly
unavailable \cite{denigPC,eidelmanPC}. We can therefore use only the
SND data, which cover the energy interval from 980 to 1380 MeV. The
statistical errors are combined with the 8\% systematic error
\cite{snd} in quadrature.

\begin{figure}
\setlength \epsfxsize{8.0cm}   
\epsffile{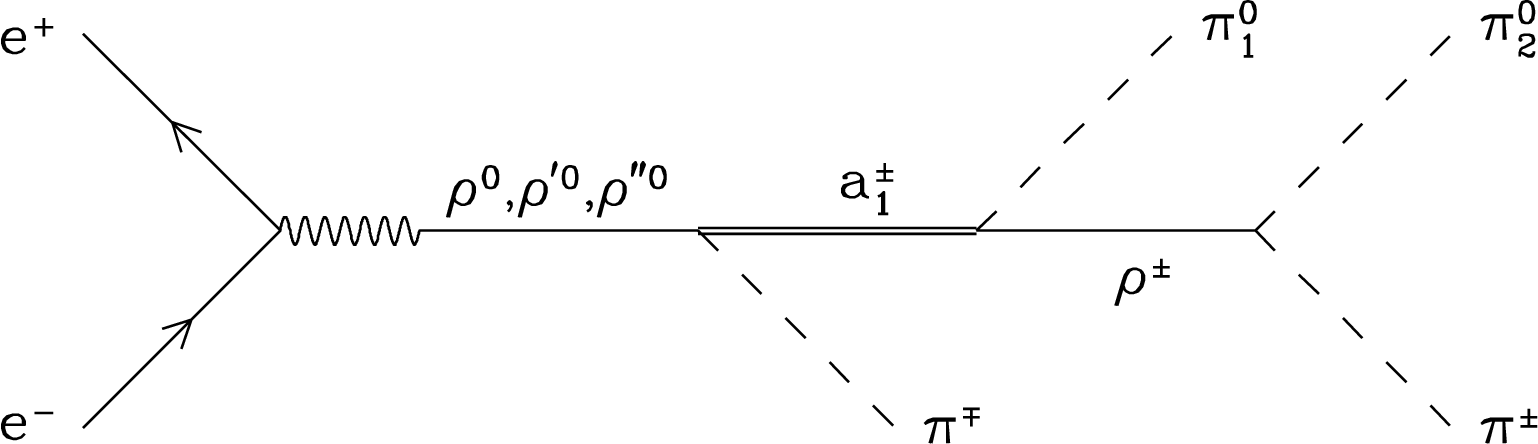}
\caption{\label{fig:ddiagmix}Two Feynman diagrams of
a pure-$a_1$ model of $\eeppppn$. Two others can be obtained by exchanging
$\pi^0$'s.}
\end{figure}

We will first compare the cross section data to a simple pure-$a_1$
model, which is characterized by four Feynman diagrams depicted in
Fig.~\ref{fig:ddiagmix}. The model is an obvious  modification of the
model used in \cite{licjur}, which is defined by a two-component $\arp$
interaction Lagrangian
\be
\label{genlag}
\lag_\arp=\frac{g_{\arp}}{\sqrt{2}}
\left(\lag_1\cos\theta+\lag_2\sin\theta\right),
\ee
where
\bea
\label{lag1}
\lag_1&=& \amu\cdot\left(\vmunu\times\partial^\nu{\fai}\right),\\
\label{lag2} \lag_2&=&
\vmunu\cdot\left(\partial^\mu\anu\times{\fai}\right),
\eea
and $\vmunu=\partial_\mu{\mathbf V}_\nu-\partial_\nu\vmu$. The
isovector composed of the $\rho$-meson field operators is denoted by
$\vmu$; similar objects for $\pi$ and $a_1$ are $\fai$ and $\amu$,
respectively. The sine of the mixing angle $\sin\theta=0.4603(28)$ was
determined in \cite{licjur} by fitting the $\eeppppc$ cross section
data from the BaBar Collaboration \cite{babar} supplemented with the
experimental value of the $D/S$ ratio in the $a_1\ra\rho\pi$ decay
\cite{DSratio}. The value of the coupling constant $g_\arp$ follows
from $\sin\theta$ and the total width of the $a_1$ meson, which was
chosen at 600~MeV.

Also defining our model is the form factor generated by the
$\rho(770)$, $\rp\equiv\rho(1450)$, and $\rpp\equiv\rho(1700)$ resonances,
\be
\label{heformfactor}
F(s)=F_\rho(s)+\delta F_\rp(s)+\epsilon F_\rpp(s).
\ee
As far as the individual contributions on the right-hand side are
concerned, we refer the reader to formulas in \cite{licjur}. Here we
only note that the complex parameters $\epsilon$ and $\delta$ as well
as the masses and widths of the $\rp$ and $\rpp$ resonances hidden in
$F_\rp$ and $F_\rpp$ were determined by fitting the four-charged-pion
BaBar data \cite{babar}.

The last ingredient of our model is connected with the structure of the
strongly interacting particles. Each vertex is usually
modified by a strong form factor to soften the interaction. In
\cite{licjur}, we used a simplified approach. We merged all form
factors to one, effective, strong form factor of the Kokoski--Isgur
\cite{kokoski} type, which multiplies the total annihilation amplitude
\be
\label{kiff}
F_{KI}(s)=\exp\left\{-\frac{s-s_0}{48\beta^2}\right\},
\ee
where $s_0=16m_\pi^2$ and $\beta=0.3695(98)$~GeV follows from the fit
to the BaBar data \cite{babar}. The same form factor is also used here.

When calculating the $\eeppppn$ excitation curve in the pure-$a_1$
model, we keep all the parameters at values determined in
\cite{licjur}. The result $\chi^2=2076$ for 35 data points is
disastrous. A poor result for the pure-$a_1$ model clearly signifies
that an additional contribution to the amplitude of the
electron-positron annihilation into two charged and two neutral pions
is needed. The intermediate states with the $\omega$ meson, considered
already by Renard in 1969 \cite{renard} and later by other authors
\cite{eidelman,achasov12,czyz,ecker34,achasov3}, are an obvious choice.

\begin{figure}
\setlength \epsfxsize{8.0cm}   
\epsffile{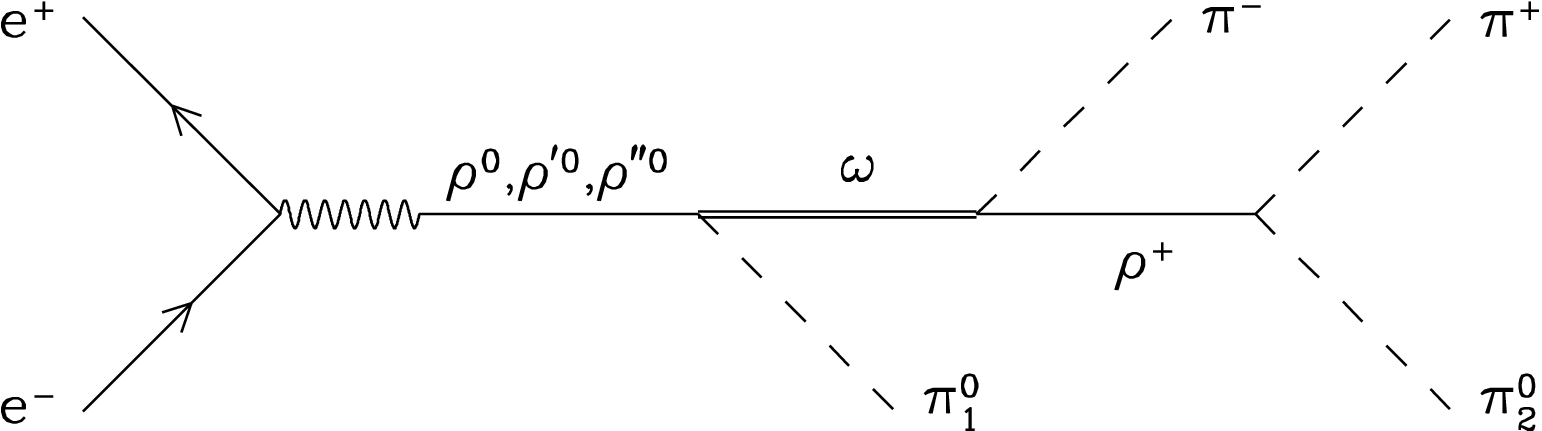} \caption{\label{fig:cdiag}The generic Feynman
diagram describing the $\omega$ contribution to $\eeppppn$. The other
five diagrams are obtained by obvious modifications.}
\end{figure}

\begin{figure}
\setlength \epsfxsize{6.95cm}   
\epsffile{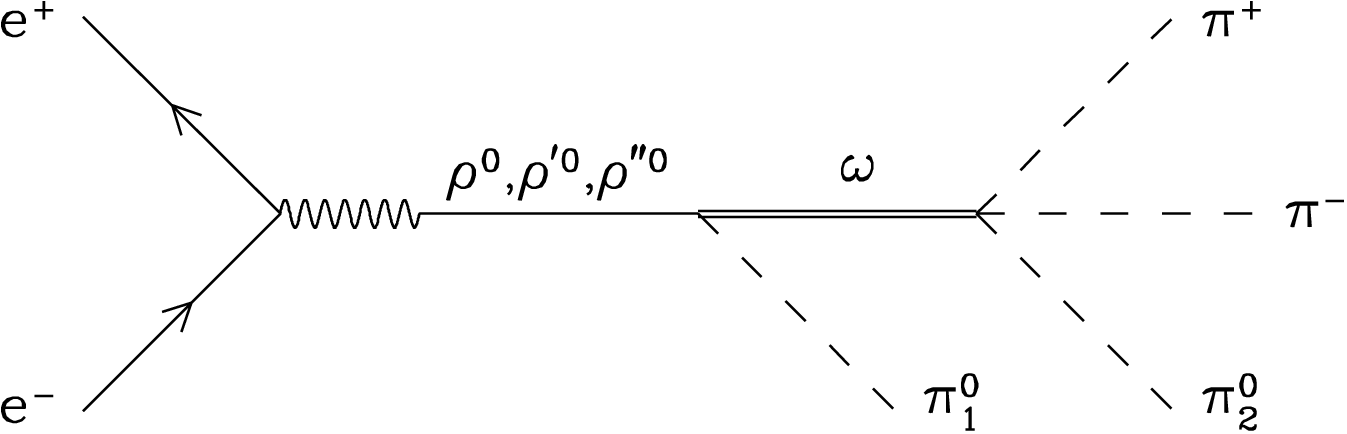} \caption{\label{fig:ccontact}One of the two Feynman
diagrams with the contact $\omega3\pi$ term. Another is obtained by
exchanging $\pi^0$'s.}
\end{figure}

We will pursue two different ways of including the intermediate states
with the $\omega$ meson. First, we adopt the approach of Eidelman,
Silagadze, and Kuraev (ESK) \cite{eidelman}, who used the anomalous
part of the chiral Lagrangian \cite{wzw,bando,kuraev} to describe the
$\orp$ and $\omega3\pi$ vertices. The Feynman diagrams are shown in
Figs.~\ref{fig:cdiag} and \ref{fig:ccontact}.
The second approach (PL) \cite{Lcecka} is more phenomenological. It
does not consider, like \cite{achasov12,achasov3}, the $\omega3\pi$
contact term. The Lagrangian
\be
\label{lagomrhpi}
{\cal L}_{\orp} =
G_\orp\epsilon_{\mu\nu\rho\sigma}
\left(\partial^\mu\omega^\nu\right)\left(\fai\cdot
\partial^\rho\mathbf{V}^\sigma\right)
\ee
has the same form as in \cite{eidelman,achasov12,achasov3}. But
now, the coupling constant $G_\orp$ is determined from the
$\omega\ra3\pi$ decay width taking into account the value of the
$\rho\pi\pi$ coupling constant $g_\rho$ as it follows from the
$\rho\ra\pi\pi$ decay width, $g_\rho^2=35.77\pm0.24$. We get
$G_\orp^2=(216.2\pm3.0)$~GeV$^{-2}$. The value of $G_\orp$ is higher
than the corresponding quantity in \cite{eidelman} by only about 2.6\%,
so the main difference between the two approaches lies in the diagrams
with the $\omega3\pi$ contact terms. To check the soundness of our
approach, we performed two tests.

First, we calculated the width of the radiative decay
$\omega\ra\pi^0\gamma$ assuming the strength of the $\rho^0\gamma$
coupling, as it follows from the normalization of the pion form factor,
\be
\label{rhogamma}
\lag_{\gamma\rho^0}=\frac{em_\rho^2}{g_\rho}A^\mu\rho^0_\mu.
\ee
The calculated branching fraction of $(9.48\pm0.28)$\% differs a little
from the current experimental value
$B(\omega\ra\pi^0\gamma)=(8.90^{+0.27}_{-0.23})\%$ \cite{pdg2006}.

Second, we determined the strength of the $\omega\gamma$ coupling from
the $\omega\ra\die$ decay width and used it in calculating the rate of
the $\pi^0\ra\gamma\gamma$ decay. The result, expressed in terms of the
$\pi^0$ mean lifetime, $\tau=(7.7\pm0.4)\times10^{-17}$~s agrees well
with the experimental value of $(8.4\pm0.6)\times10^{-17}$~s.

What remains unsettled is a possible transfer-momentum-squared ($t$)
dependence of the $\rho\gamma$ coupling. In fact, the experimental
width of the $\rho^0\ra\die$ decay, where $t=m_\rho^2$, requires about
20\% stronger coupling than that indicated in \rf{rhogamma}. The latter
gives good results for the $t=0$ processes $\omega\ra\pi^0\gamma$ and
$\pi^0\ra\gamma\gamma$. Our derivation \cite{licjur} of the
$\die\ra4\pi$ cross section formula used the standard $\gamma\rho^0$
coupling \rf{rhogamma} and assumed that all the $t$ dependence is
absorbed in the form factor. We use the same approach here.

Now, we can also vary the form-factor parameters $\delta$ and
$\epsilon$, as the structure of the intermediate states is different
from the pure-$a_1$ model. We will distinguish them from the $\pmpm$
case by primes. There is no free parameter connected with the
$\omega\pi$ intermediate states. The results are shown in
Table~\ref{tab:a1omega}.
\begin{table}
\caption{\label{tab:a1omega} Fitting the $a_1$+$\omega$ model to the
$\eeppppn$ cross section data ($35$ data points).}
\begin{ruledtabular}
\begin{tabular}{ccc}
Approach to $\omega$ & ESK \cite{eidelman} & PL \cite{Lcecka}     \\
\hline    
 $\chi^2/$NDF        & 2.19   & 0.82  \\
 CL (\%)             & 0.01   & 74.9  \\
 Re~$\dep$  &-0.52(25)  &-0.36(27) \\
 Im~$\dep$  &-1.27(25)  &-1.05(32) \\
 Re~$\ep$   &-0.79(34)  &-0.71(37) \\
 Im~$\ep$   & 0.953(97) & 0.628(79) \\
\end{tabular}
\end{ruledtabular}
\end{table}

\begin{figure}
\setlength \epsfxsize{8.6cm} \epsffile{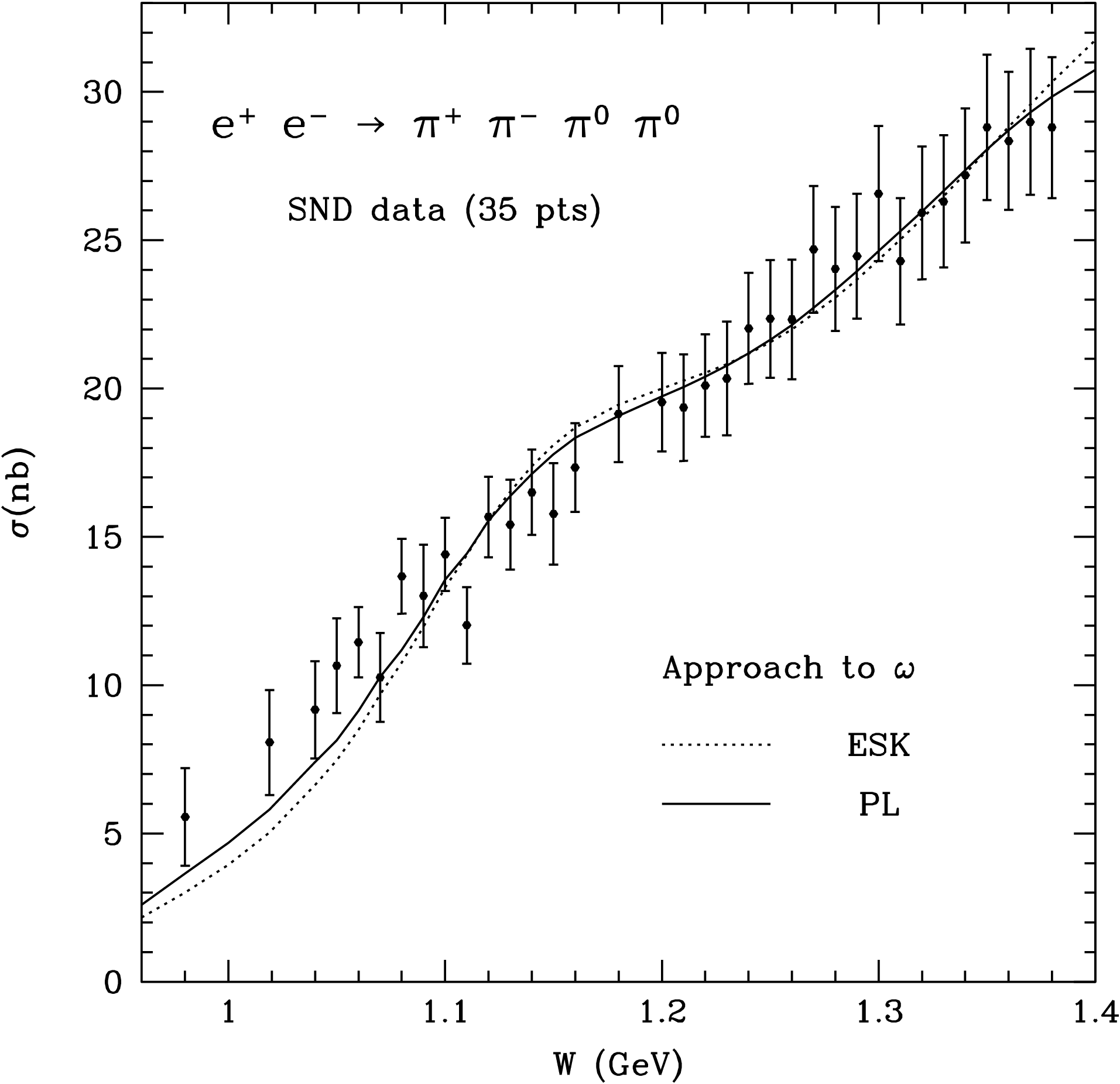}
\caption{\label{fig:cde}Comparison of the $a_1$+$\omega$+$h_1$ model
with the $\eeppppn$ data for two approaches to $\omega$.}
\end{figure}

In an effort to further improve the agreement of our model with data,
we include the intermediate states with the isoscalar axial-vector
meson $h_1(1170)$. The corresponding Feynman diagrams can be obtained from
those in Fig.~\ref{fig:cdiag} by replacing $\omega$ with $h_1$. The
interaction Lagrangian is again assumed in a two-component form similar to
\rf{genlag} but respecting the isoscalar character of $h_1(1170)$,
\be
\label{h1lag}
\lag_\hrp=\frac{g_{\hrp}}{\sqrt{3}}
\left(\lag_a\cos\eta+\lag_b\sin\eta\right),
\ee
where
\bea
\label{laga}
\lag_a&=& h^\mu \cdot\left(\vmunu\cdot\partial^\nu{\fai}\right),\\
\label{lagb}
\lag_b&=& -\partial^\mu{h^\nu} \cdot\left(\vmunu\cdot\fai\right).
\eea
In the following, the sine of the mixing angle $\eta$ will be varied to
achieve the best possible description of the cross section data.
For each $\sin\eta$ the coupling constant will be determined from the
condition that the total width of the $h_1(1170)$, calculated as
$\Gamma(h_1\ra\rho\pi)$, should be equal to 360~MeV. While fitting the
$\eeppppn$ data the mixing angle of the $\arp$ Lagrangian \rf{genlag}
is kept fixed at the value determined in the $\pmpm$ case, as it
represents a universal process-independent parameter. The results of
the fit are shown in Table~\ref{tab:a1omegah1} for both approaches to
the intermediate states with $\omega$. It is clear that the inclusion
of the $h_1\pi$ intermediate states greatly improves the confidence
level of the model with $\omega$ described by the ESK scheme. The
confidence level of the model utilizing the PL scheme for $\omega$ also
rises, but because it was already high in the $a_1$+$\omega$ model, the
inclusion of the intermediate states with $h_1$ is not necessary. The
excitation curves are compared to data in Fig.~\ref{fig:cde}.

\begin{table}
\caption{\label{tab:a1omegah1}Fitting the $a_1$+$\omega$+$h_1$ model to
the $\eeppppn$ cross section data ($35$ data points).}
\begin{ruledtabular}
\begin{tabular}{ccc}
Approach to $\omega$ & ESK \cite{eidelman}  &  PL \cite{Lcecka} \\
\hline    
 $\chi^2/$NDF   & 1.18        & 0.80       \\
 CL (\%)        & 22.4        & 77.8       \\
 $\sin\eta$     & 0.3434(36)  & 0.3433(46) \\
 Re~$\dep$   & 0.092(62)   & 0.102(74) \\
 Im~$\dep$   & 0.028(22)   & 0.035(23) \\
 Re~$\ep$    & 0.022(71)   & 0.028(86) \\
 Im~$\ep$    &-0.030(58)   &-0.049(65) \\
\end{tabular}
\end{ruledtabular}
\end{table}

Following the idea that a simultaneous fit to more processes may lead
to a more precise value of the mixing angle of the $\arp$ Lagrangian, we
are now going to perform a joint fit to the $\eeppppc$ and $\eeppppn$
cross section data. We merge the pure-$a_1$ model of our previous paper
\cite{licjur} with the $a_1$+$\omega$+$h_1$ model described here. For
the $\omega\pi$ intermediate states we will use the PL version, which
describes the data better than ESK. Simultaneous handling of
both four-pion channels enables us to use a more correct description of
the $\rho(770)$ part of the electromagnetic form factor
\rf{heformfactor}, namely,
\be \label{frho}
F_\rho(s)=\frac{M_\rho^2(0)}{M_\rho^2(s)-s-im_\rho\tilde\Gamma_\rho(s)},
\ee
where $M_\rho(s)$ is the running mass of the $\rho$ meson
calculated in \cite{ratio}. The total decay width of the $\rho^0$ meson,
\be
\tilde\Gamma_\rho(s)=\Gamma_\rho(s)+\Gamma_{\pmpm}(s)+\Gamma_{\pmnn}(s)
\ee
now includes not only the contribution of several two- and
three-body decay channels $\Gamma_\rho(s)$ given in \cite{ratio}, but
also the contributions from both four-pion decay modes.
Similarly to \cite{licjur} we also consider the masses and widths of
$\rp$ and $\rpp$ as free parameters. They have the same values in both
channels, as well as the $\arp$  Lagrangian mixing parameter
$\sin\theta$ and the parameter $\beta$ of the strong form factor
\rf{kiff}. Besides those six common parameters, there are two sets of
parameters specific for each of the two four-pion annihilation
channels. The charged-pion-channel set contains four real parameters
entering the electromagnetic form factor \rf{heformfactor}. The four
form-factor parameters in the mixed-pion channel are distinguished from
those in the $\pmpm$ channel by primes. The fifth parameter in the
mixed-pion channel is the $\hrp$ Lagrangian mixing parameter
$\sin\eta$. All together, this makes 15 real free parameters.

In Table~\ref{tab2} we present the optimized values of all free
parameters. The corresponding confidence level is 28.4~\%. If we
compare them with those obtained in individual models
(Table~\ref{tab:a1omegah1} here and Table~VI in \cite{licjur}), we can
say that they are in good agreement. Only the error domains of
$m_{\rp}$, $m_{\rpp}$, and $\delta$ do not overlap, but the
disagreement is very small. The excitation curves of individual
reactions do not visually differ from those obtained when fitting the
two models separately and are not shown.
\begin{table}
\caption{\label{tab2}Joint fit of the pure-$a_1$ model of $\eeppppc$
and the $a_1$+$\omega$(PL)+$h_1$ model of $\eeppppn$ to a combined set
of the cross section data and the $D/S$ ratio ($180$ data points).}
\begin{ruledtabular}
\begin{tabular}{cc|cc}
 Quantity   & Value  & Quantity   & Value \\
\hline
 $\chi^2/$NDF       & 1.06      & Re $\dep$       & 0.088(73)  \\
 $m_{\rp}$ (GeV)    & 1.383(16) & Im $\dep$       & 0.024(22)  \\
 $\Gamma_{\rp}$ (GeV)& 0.551(21)& Re $\ep$        & 0.025(81)  \\
 $m_{\rpp}$ (GeV) & 1.883(18)   & Im $\ep$        & -0.011(74) \\
 $\Gamma_{\rpp}$ (GeV)& 0.237(35) & Re $\delta$   & 0.171(14)  \\
 $\sin\theta$   & 0.4662(52)    & Im $\delta$     & 0.045(29)  \\
 $\beta$ (GeV)  & 0.3617(70)    & Re $\epsilon$   & 0.0002(14) \\
 $\sin\eta$     & 0.336(11)     & Im $\epsilon$   &-0.0050(12) \\
\end{tabular}
\end{ruledtabular}
\end{table}

Unfortunately, our goal to narrow the interval of the $\arp$ Lagrangian
mixing parameter and thus make the calculations of the dilepton and
photon production from hadron gas more reliable (see the analysis in
\cite{gaogale}) has not been reached. The uncertainty of $\sin\theta$
is larger than that obtained in \cite{licjur}. We hope that the
situation will improve when more precise cross section data in the
mixed-pion channel are available. Identifying the essential
contributions to the electron-positron annihilation into the four-pion
final states is important for the reliable assessment of the dilepton
production by the four-pion annihilation in hadron gas \cite{joerg}. A
new result of our work is the mixing angle of the $\hrp$ Lagrangian
$\sin\eta\approx0.34$. It may help in investigating the role of the
$h_1(1170)$ resonance in thermal production of dileptons and photons
from hadron gas, which has been ignored so far.

We thank T.~Barnes, A.~Denig, and S.~I.~Eidelman for useful
correspondence. This work was supported by the Czech Ministry of
Education, Youth and Sports under Contract No. MSM6840770029, No.
MSM4781305903, and No. LC07050.

\end{document}